# Simultaneous amplitude and phase modulation for wide-field nonlinear microscopy applications

Miguel Carbonell-Leal[1], Gladys Mínguez-Vega[1], Jesús Lancis[1] & Omel Mendoza-Yero[1*]

[1]Institut de Noves Tecnologies de la Imatge (INIT), Universitat Jaume I, 12080 Castelló, Spain.
*Corresponding author: omendoza@fca.uji.es

In wide-field nonlinear microscopy, wavefront modulation by means of phase-only spatial light modulators (SLMs) allows achieving simultaneous two-photon excitation and fluorescence emission from specific region-of-interests (ROIs) of biological specimens. This is basically accomplished at the illumination path of the microscope by the reconstruction of computer generated holograms (CGHs) onto the sample plane. However, as two-photon absorption (TPA) is inherently an intensity-square dependent process and iterative Fourier transform algorithms (IFTAs) can only approximate the illumination of selected ROIs with the reconstructed CGHs, both signal acquisition and/or image formation can be largely affected by the spatial irregularities of the illumination patterns. In addition, the speckle associated with the superposition of coherent light at the selected ROIs prevents illumination strategies based on CGHs to be successfully used for large-area (more than 50x50 μm$^2$) excitation tasks. To overcome these limitations, we propose an alternative complex illumination method (CIM) able to generate simultaneous nonlinear excitation of large-area ROIs with full control over the amplitude and phase of the optical wavefront. We experimentally demonstrate spatially uniform illumination, as well as structured illumination with user-defined intensity levels onto micrometric but large-area ROIs. Furthermore, a proof-of-concept experiment on wide-field second harmonic generation (SHG) is provided. We believe that the proposed CIM could find applications in wide-field nonlinear microscopy, particularly for speed up signal acquisition time or improve two-photon image formation.

## 1. Introduction

In some recent introduced multiphoton microscopy techniques, simultaneous excitation and signal collection from multiple specific cell populations have become into key tools for monitoring the cellular activity [1–3]. This basically happens because multiple millisecond time response signals e.g., fluorescence-lifetime signals, produced by many cellular ensembles cannot be acquired by means of conventional scanning methods. In point by point scanning methods, the temporal resolution is limited by the signal-to noise ratio, which is close-related to the pixel dwell time. That is, the longer the pixel dwell time, the higher the signal-to-noise ratio but consequently, temporal resolution gets worse. Additionally, it is apparent that by increasing the laser intensity one can also increase signal-to-noise ratio, however unwanted photo-induced biological damage due to single or multiphoton light absorption may also appear (approximately for average power beyond 10 mW).

An alternative illumination technique employed for speed up acquisition time in multiphoton microscopy relies on the parallel excitation of the cellular targets by using multifocal irradiance patterns. With this technique, the acquisition time can be reduced by at least the number of excitation foci. On this topic, multifocal multiphoton microscopy can be carried out by several methods including, but not limited to, the use of microlens arrays [4,5], beam splitting [6], or Fresnel holograms [7]. However, it is apparent that, multifocal illumination allows us to excite only focal-size sites in the cells, which certainty limit the acquired information e.g., voltage signals, to those focal regions determined by the spatial features of the excitation foci.

On the other hand, the application of wide-field illumination techniques together with the utilization of high-sensitivity cameras (e.g., TE-cooled, ultra-sensitive photon detecting, electron multiplying charge-coupled camera) can be regarded as a prominent solution to simultaneously excite and collect not just at focal-size sites in the cells, but also from large area of them. In principle, all the field-of-view under the numerical aperture of the objective could be turned into single-photon fluorescence signal, although it is less likely to induce multiphoton absorption and fluorescence emission in biological cells due to both the exponential intensity requirements of nonlinear processes and the spreading of the laser power over the extended regions. Hence, instead of conventional mode locked Ti: sapphire laser oscillators, additional chirped pulse amplification stages with average power capabilities up to several watts and repetition rates of tens of kHz become better candidates of choice for performing wide-field multiphoton microscopy [8,9].

In this context, commercially available SLMs have been employed to manipulate the optical wavefront of ultrashort pulses in order to achieve simultaneous excitation of ROIs within a cellular ensemble [2, 10–15]. On this topic, among the extensive variety of reported illumination methods, those based on the use of phase-only SLMs stand out because of their potential to manage light with relatively high throughput. In particular, wide-field nonlinear excitation has been experimentally demonstrated by the coherent reconstruction of CGHs at the sample plane. In this case, although long part of the available energy can be transferred to specific ROIs within the cellular ensembles, there still exist several problems to face out. For instance, unwanted effects like angular dispersion or chromatic aberrations may cause spatiotemporal distortions of the reconstructed hologram, which clearly decrease the signal-to-noise ratio and harm the image acquisition process. In addition, the coherent noise (speckle) associated with the superposition of coherent light with different path lengths prevents wide-field excitation based on CGHs to be well-accomplished over other ROIs that are not the typical line-shaped irradiance patterns reported in the literature [7,12]. Therefore, under coherent illumination, the reconstruction of CGHs is currently unable to generate spatially homogeneous irradiance patterns over ROIs with dimensions in the order of the typical cellular-sizes (about 50x50 μm²), not to mention its inefficiency to precisely manipulate the amount of energy put into then e.g., to generate multiple-intensity level illumination patterns. In this manuscript we propose an interferometric CIM to get wide-field nonlinear excitation of cellular-size ROIs with full and independent control over the amplitude and phase of the illumination source. Its ability to precisely modify the complex field of the spatially coherent light emitted by our femtosecond laser is experimentally shown to be crucial for obtaining not only homogenous irradiance patterns (without speckle) over above-mentioned ROIs, but also multiple-intensity levels. To validate our method, we experimentally demonstrate a proof-of-concept experiment addressed to induce wide-field SHG in a Type 1 $\beta-BaB_2O_4$ (BBO) crystal. All optical manage of the illumination patterns is carried out with a single phase element implemented onto a phase-only SLM. The phase element is computer generated by using a complex field encoding method based on double-phase hologram theory that was previously reported in Ref [15]. Some results achieved with our CIM were compared with similar ones obtained with the reconstruction of CGHs calculated by IFTAs.

## 2. Complex illumination method

In this section we introduce the theory underlying the CIM. Its optical setup is basically composed of two consecutive optical modules that we refer throughout the whole manuscript as complex field encoding module (CFEM) and optical demagnifying module (ODM), see Fig 1.

In the CFEM the incident optical wavefront is phase-only modulated to get user-defined complex fields at the output plane. In accordance with the theory [15], any two-dimensional complex field expressed in the form $U(x,y) = A(x,y)e^{i\varphi(x,y)}$, where $A(x,y)$ and $\varphi(x,y)$ hold for the amplitude and phase functions, respectively, can be conveniently rewritten as:

$$U(x,y) = Be^{i\theta(x,y)} + Be^{i\vartheta(x,y)} \qquad (1)$$

In Eq. (1), $B = A_{max}/2$ is an amplitude constant term (it does not depends on the transversal coordinates $x, y$), $A_{max}$ is the maximum value of $A(x,y)$, and the new phase functions $\theta(x,y)$ and $\vartheta(x,y)$ can be calculated from $A(x,y)$ and $\varphi(x,y)$ by using the expressions:

$$\theta(x,y) = \varphi(x,y) + \cos^{-1}[A(x,y)/A_{max}] \qquad (2)$$
$$\vartheta(x,y) = \varphi(x,y) - \cos^{-1}[A(x,y)/A_{max}] \qquad (3)$$

From a physical point of view (for $A_{max} = 2$), equations (1-3) mean that the complex field $U(x,y)$ can be retrieved from the coherent interference of the two uniform waves $e^{i\theta(x,y)}$ and $e^{i\vartheta(x,y)}$. Such interference can be carried out by means of a common-path interferometer made up of a spatially filtered 4f optical system. At the input plane of this optical system is placed the screen of a phase-only SLM where the following phase element $\alpha(x,y)$ is implemented

$$\alpha(x,y) = M_1(x,y)\theta(x,y) + M_2(x,y)\vartheta(x,y) \qquad (4)$$

This phase element $\alpha(x,y)$ is determined from the transversal spatial mapping of $e^{i\theta(x,y)}$ and $e^{i\vartheta(x,y)}$ with two-dimensional binary gratings $M_1(x,y)$ and $M_2(x,y)$ (checkerboard patterns) taken at the Nyquist limit, that is

$$M_1(x,y)e^{i\theta(x,y)} + M_2(x,y)e^{i\vartheta(x,y)} = e^{i\alpha(x,y)} \qquad (5)$$

The checkerboard patterns fulfill the complementary condition $M_1(x,y) + M_2(x,y) = 1$. Here, it should be noted that send the phase element $\alpha(x,y)$ to the SLM does not guaranties the interference of uniform waves to happen. This is done by means of a suited spatial filter placed at the Fourier plane of the imaging system. It can be shown that if we use a filter to block all diffraction orders but the zero one, we are able to exactly retrieve the full spectrum of the original complex field in the Fourier plane [15]. Consequently, at the output plane we find the convolution of the magnified spatially reversed complex field with the Fourier transform of the filter. It means that full amplitude and phase information of the complex field can be retrieved at the output plane, except for certain loss of spatial resolution due to the convolution operation. Therefore, the CFEM consists of a common-path interferometer that coherently sums two uniform waves spatially mapped and encoded into a phase-only SLM.

On the other hand, the ODM allows us to proper rescale the complex field generated at the output plane of the CFEM to the sample plane. This last optical setup is composed of a refractive achromatic lens with long-distance focal length and an infinity corrected microscope objective, which generate a demagnified image of the reconstructed complex field onto the sample plane. The whole optical setup used to demonstrate the CIM is shown in Fig. 1, and it is described in details in next section.

## 3. Complex illumination patterns

To demonstrate the ability of the method to generate user-defined complex illumination patterns at micrometric scale we use the experimental setup shown in Fig. 1. The ultrashort pulses emitted by a Ti: Sapphire amplifier laser (30 fs amplitude full-width at half maximum (FWHM), central wavelength 800 nm, bandwidth of 50 nm FWHM, maximum energy per pulse of 0.8 mJ, and 1 kHz repetition rate) are employed as a light source. The SLM (HOLOEYE PLUTO, resolution 1920x1080, and pixel pitch 8 μm) used in our experiments was previously calibrated [16] for a significant set of frequency components of the light pulse with the help of three bandpass filters of 10 nm FWHM. The similar phase responses (calibration curves) obtained for the three different spectral lines of the ultrashort pulse allows performing wavefront modulation of ultrashort pulses with a single phase element $\alpha(x,y)$. In addition, to mitigate the effect of abrupt spatial variations in the phase modulation [17], the two-dimensional binary gratings $M_1(x,y)$ and $M_2(x,y)$ were constructed with 4x4 pixel-cells.

In the optical setup, the light is sent to the SLM by using silver mirrors and reflected from it forming a small angle (about 3 degrees) with respect to the direction of normal incidence. In these conditions, the desired complex field $U(x, y)$ is reconstructed at the output plane of the CFEM. The CFEM is made up of a couple refractive lenses (L$_1$ and L$_2$) with focal lengths L$_1$ = 1000 mm and L$_2$ = 500 mm, which result in a transversal magnification of ½ at its output plane. In addition, a circular iris is placed at the Fourier plane to filter all diffraction orders but the zero one. Finally, the ODM allows complex field illumination $U(x, y)$ to be imaged into the sample plane.

A rough coupling between the size of the illumination pattern and that of the sample is realized with a proper selection of optical components of the ODM. In our experiment, it is made up of an achromatic lens (L$_3$) of focal length 400 mm and a 20x infinity-corrected microscope objective (MO$_1$), which led to a total demagnification of about 1/73 at the sample plane. The transmitted light is then collected with the help of a 50x infinity-corrected microscope objective (MO$_2$) and sent to a CCD camera (Basler avA1600-50gm), where the image of the sample is recorded (see Fig. 1). To adjust the intensity of the light at the sample plane a set broadband neutral filters was employed. A more accurate selection of the illumination zone over the sample is done later once the specific ROIs are defined. At this point it is possible to establish the main features of the illumination pattern $U(x, y)$. In particular, its amplitude function $A(x, y)$, for instance: a multilevel spatial energy distribution, or its phase function $\varphi(x, y)$, for instance: a proper phase distribution to compensate for specimen-induced or microscope objective spherical aberrations.

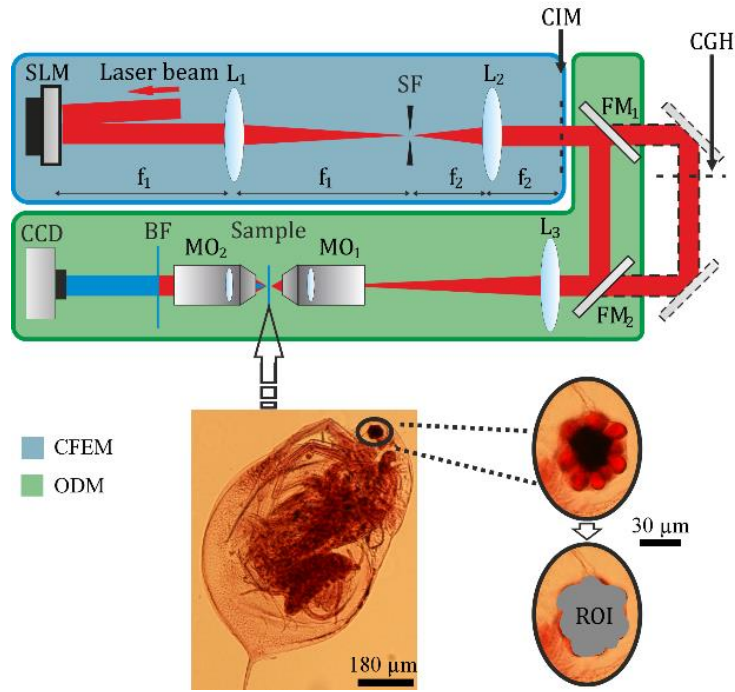

**Figure 1.** Optical setup. Complex illumination method (CIM) implemented by means of two main modules: complex field encoding module (CFEM), and optical demagnifying module (ODM).

In Fig. 2, an example corresponding to a complex illumination pattern generated at the sample plane is theoretically and experimentally shown. In this case, we simulate a ROI within a Daphnia specimen (water flea) that coincides with its eye. Note that, such ROI has approximate dimensions of 60x60 µm$^2$, which is about the order of most cell sizes. Specifically, the encoded complex field given in top part of Fig. 2 is characterized by an amplitude pattern having four different energy levels distributed among quadrants within the eye, whereas its spatial phase is given by the analytical function $\varphi(x, y) = \pi/2 \sin[2\pi x/(5\lambda)]\cos[2\pi y/(5\lambda)]$. From the above-defined amplitude and phase parameters we can calculate the phase element $\alpha(x, y)$ used to reconstruct the expected complex field. To do that, we follow the procedure described in section 2, also represented schematically in central part of Fig. 2. Here, to clearly see the roll of sampling gratings $M_1(x, y)$ and $M_2(x, y)$ on this procedure, the phase element $\alpha(x, y)$ was represented with pixel-cells of 60x60.

Polarization-based phase shifting technique [18] is applied to measure the amplitude and phase of the complex illumination pattern at the sample plane. Please, note again that the optical setup shown in Fig. 1 can be regarded as a common-path interferometer which perform the coherent interference of two spatially sampled uniform waves, $e^{i\theta(x,y)}$ and $e^{i\vartheta(x,y)}$. So, this phase shifting technique can be accomplished by simple addition of a couple of polarization-dependent optical elements e.g., two polarizers. Specifically, a broadband halfwave plate (EKSMA OPTICS 460-4215) was placed before the SLM plane to rotate the original direction of polarization of all frequency components of the ultrashort pulse 45 degrees with respect to the SLM director orientation. This allows only part of the incident light to be phase modulated (object beam), while the remaining one (reference beam) is not diffracted. In order to generate four different phase shifting interferograms at the sample plane, uniform phases with steps of $\pi/2$ radians are added to the phase element $\alpha(x, y)$. The interferograms are formed after recombining both reference and object beams with the help of a broadband linear polarizer (EKSMA OPTICS 420 0526M), this time located before the 20x microscope objective MO$_1$. In practice, the position of the rotational angle with respect to SLM director orientation determines the amount of diffracted and non-diffracted light within the interferograms. In the bottom-central part of Fig. 2, the interferograms measured for the phase steps 0, π/2, π, and 3π/2 are shown. These interferograms were recorded without moving the position of the camera in the optical setup. The experimental amplitude and phase patterns determined from the above phase shifting interferograms are given in the bottom part of Fig. 2. It is apparent that experimental results are in good agreement with the theory. However, as it might be expected due to the convolution operation (see section 2) and the used pixel-cell, the spatial resolution of recorded amplitude and phase patterns is a few pixels lower than corresponding theoretical ones. Even so, the root-mean-square error (RMSE) between theory and experiment yields discrepancies of 6.8% and 14% for amplitude and phase images, respectively.

Now, we test the usefulness of the CIM to synthetize amplitude-only irradiance patterns at micrometric scale. On one side, we are interested in the generation of spatially uniform irradiance patterns for simultaneously wide-field illumination. Note that the excitation of extended ROIs with non-uniform irradiance patterns may cause image blurring due to different signal responses of similar biological components. In wide-field two-photon microscopy this is crucial because the recorded signal depends on the square of the light intensity. In other cases, beam shaping of coherent radiation for simultaneous excitation of extended ROIs (not by multifocal excitation) cannot be accomplished without coherent noise (speckle), which clearly affect the quality of the acquired images.

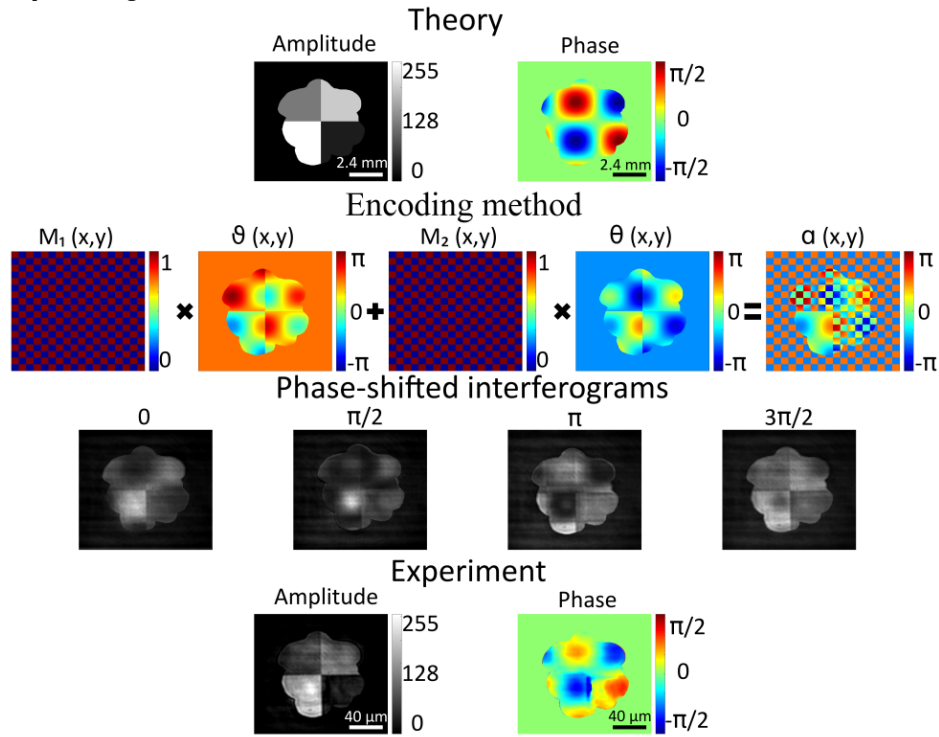

**Figure 2.** Generic example: measured amplitude and phase of a given complex field at the SLM (top) and sample (bottom) planes. Related CIM images and measured interferograms are also shown (central part). Scales are included as insets at the right-bottom part of e each irradiance pattern.

In contrast, by applying the proposed CIM we obtain spatial uniform illumination of extended ROIs without speckle (please, see top-right part of Fig. 3). In this experiment, as before, we select a ROI whose shape is determined by the borders of the Daphnia 'eye. The first column of Fig. 3 shows the phase elements $\alpha(x, y)$ represented with a pixel-cell of 6x6. Here, the RMSE between theory and experiment yields discrepancies of 7%. The spatial uniformity of the recorded pattern mainly depends on the spatial profile of the laser beam at the SLM plane, and also to a lesser extent of optical irregularities (scratches in mirrors or lenses, unwanted dust on the surface of optical components, etc.). Furthermore, as this irradiance pattern is basically constructed by conjugating the SLM plane with the sample plane, measured irradiance patterns are free from coherent noise.

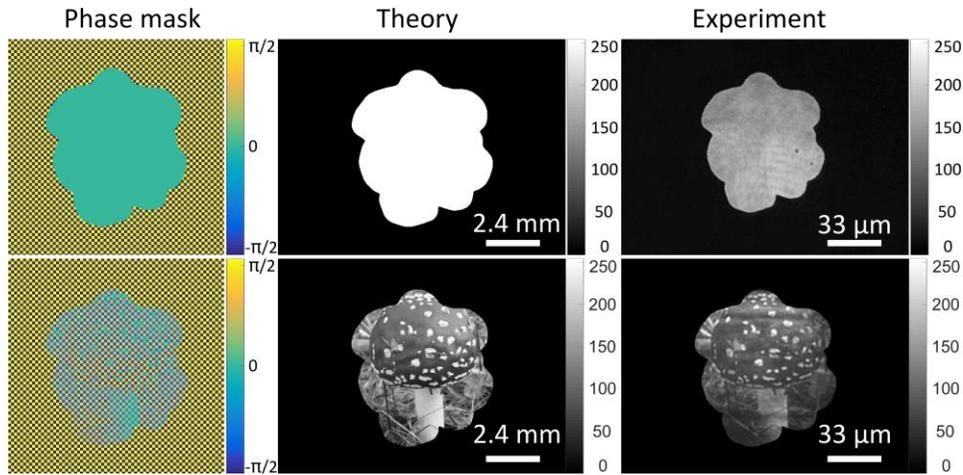

**Figure 3.** Theoretical and experimental amplitude-only irradiance patterns at the SLM and sample planes, respectively. Scales are included as insets at the right-bottom part of each irradiance pattern. In the first column, corresponding phase elements, represented with a pixel cell of 6x6, are also shown.

On the other side, we push the CIM to the limit by implementing irradiance patterns with multiple energy levels and complex spatial structure. To this end, we reproduce inside of the ROI given by the Daphnia's eye (about 60x60 $\mu m^2$ only) part of a mushroom image. After a visual comparison of results shown in Fig. 3, one can realize that the recorded irradiance pattern reproduces very well the expected one. In fact, we are able to spatially resolve image details in the sample plane of the order of 5 $\mu m$ i.e., the white spots. In this case, the RMSE between theory and experiment yields discrepancies of 10%. With this experiment, we want to show the potential of the CIM to manage different energy levels of excitation simultaneously. It is apparent that different constituents of cells and/or fluorophores might imply different excitation thresholds. So, in principle, one can think in setting energy-custom illumination patterns depending on the nature of the biological sample. This fine control over the energy content of the illumination pattern can be also useful to adjust the energy levels over the whole sample without the necessity of using conventional neutral filters.

## 4. Non-linear excitation experiment

In this section we experimentally demonstrate wide-field nonlinear excitation phenomena at micrometric scale. In particular, we show that our CIM can generate simultaneous and controllable second harmonic signal within a selected ROI. The chosen complex illumination pattern is made up of a uniform amplitude, and a proper phase able to compensate few optical aberrations found at the sample plane. This phase was determined as the complementary of that one obtained with a Shack–Hartmann wavefront placed at the same plane of the camera with the help of an additional beam splitter (not shown in Fig. 1), when a matrix of zero radians is sent to the SLM. The ROI resembles again the structure of the eye of a Daphnia (similar to that shown in top part of Fig. 3). In addition, to compare the proposed CIM with a well-established illumination method, we include in the optical setup a delay line for synthetizing similar amplitude illumination patterns, but this time by means of CGHs. For this purpose the optical setup is designed such that the reconstructed CGH appears at the same distance from the lens $L_3$ as the complex field $U(x, y)$ does. The CGHs were designed following the Gerchberg–Saxton IFTA [19] developed in two different stages. In the first stage we perform iterations by applying only phase freedom, while the second stage the phase quantization of the hologram is softly restricted until the desired number of phase levels is reached. The algorithm converges when the result of the iteration improves the RMSE by an amount of $10^{-8}$, compared with the previous one. In the experiment, we put a BBO crystal with dimensions 10 mm × 10 mm × 0.02 mm at the sample plane, as shown in Fig. 1. The unconverted infrared wavelengths of the ultrashort laser pulse were conveniently filtered with a BG39-Schott crystal before the CCD camera.

At this point, we focus our attention on irradiance rather than phase of the encoded complex field. The main reason is that, for our experimental conditions, the employed aberration-compensation phase was almost flat. Hence, in the top part of Fig. 4, we show only the experimental irradiance patterns reconstructed at the sample plane with the two different illumination methods. These patterns were recorded with the camera located at the position shown in Fig. 1, before placing the BBO crystal at the sample plane. From the visual comparison of both images, differences among them are apparent. The reconstructed CGH roughly approximates the expected uniform irradiance pattern. Its failed attempt to reproduce a uniform illumination pattern is limited to a clear reinforcement of light at the borders of the ROI, together with the presence of coherent noise. In contrast, the irradiance pattern obtained with our CIM is almost fully flat. To better illustrate the differences between both patterns we show the irradiance profile along the central part of the ROIs. The calculated fluctuations with respect to the ideal flat profile are 62% for the CGH method, against 12% for the CIM.

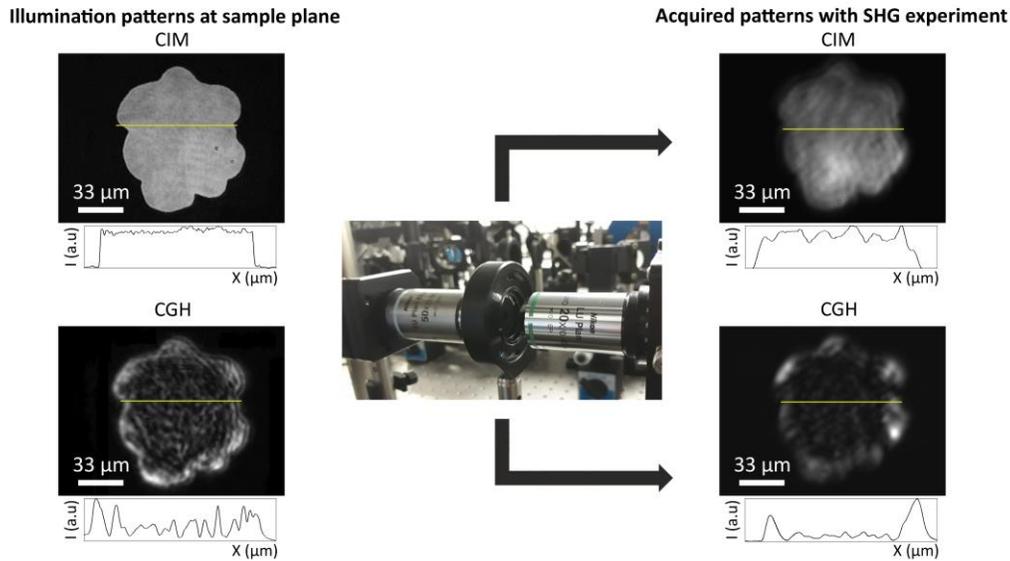

**Figure 4.** Experimental images of wide-field second harmonic emission from a Daphnia 'eye ROI after employed both CGHs and our illumination methods. Scales and central irradiance profiles are included as insets at the left-bottom and right-bottom parts of each irradiance pattern, respectively.

Once the nonlinear medium is placed at the focal plane, we are able to measure second harmonic signal. The corresponding recorded normalized images are shown now in the bottom part of Fig. 4. As it might be expected, the nonlinear response recorded in this experiment is highly dependent on the degree of flatness of the illumination patterns. That is, second harmonic image due to the illumination with reconstructed CGHs shows almost no light at those sites different from the edges of the ROI. For completeness, we include at the bottom part of each image a spatial profile of these patterns, taken again at the center of each one. This time the fluctuations with respect to the ideal flat profile are 77% for the CGH method, and 23% for the CIM.

From these experimental results, one can realize that, in principle, wide-field second harmonic excitation of biological samples is feasible under uniform and proper illumination conditions. On this respect, the introduced CIM plays a fundamental role in the management of the complex field of laser beams at the sample plane. We believe that parallel excitation of extended ROIs under controllable illumination conditions could positively contribute to decrease the acquisition time, and consequently faster the image formation in some real non-linear microscopy applications.

## 5. General considerations

In this section we will discuss on significant practical issues of the CIM. For instance, in order to highly mitigate the effect of non-diffractive light (zero order) onto the sample plane, the light in the ODM enters the microscope objective $MO_1$ under low-focusing, instead of typical plane-wave configuration. In this way, non-diffracted light exits the microscope objective parallel to the optical axis without forming any potentially-dangerous focus near the output plane of the 4f optical system (sample plane). Hence, at the sample plane the excitation pattern is mainly determined by the useful light diffracted at the SLM and not by the unwanted zero-order contribution. As the fill factor of commercially available SLMs is usually very high (93% for our SLM), the signal-to-noise ratio determined by diffracted and non-diffracted light is really good. In addition, thanks to the use of a long-distance focal length achromatic lens to direct the light into the microscope objective $MO_1$, the effects of spherical aberration at the sample plane can be disregarded. Perhaps, this can be better understood in terms of the produced Rayleigh range inside the microscope objective, which is longer and longer as far as we increase the focal length of above-mentioned achromatic lens.

Now we analyze the efficiency of the CIM. In general the efficiency of a method based on encoding information onto a SLM depends on several factors, including but not limited to technical specifications of the SLM, the type of information (amplitude or phase) that one needs to encode, characteristics of the light source, and/or the orientation of the SLM with respect to the incident light (oblique or straight light incident). In the proposed illumination method main energy losses come from the diffraction of light at the phase element, as well as due to the physical specifications of our SLM (in accordance with the manufacturer: 80% diffraction efficiency, 65-95% reflectivity, 93% fill factor, so the total light efficiency can be more than 50%). Another important issue to consider here is the size of the encoding patterns, which is directly proportional to the efficiency of the method. In other words, as in our illumination method the SLM plane is (except for the spatial filtering process) conjugated to the sample plane, an increase in the area of the encoded pattern at the SLM leads to a more available light for illumination purposes. On this context, we measured the efficiency of the illumination method for the generation of irradiance patterns with the spatial shape of circumferences having different radii, and compared these results with similar ones obtained from the reconstruction of CGHs with equal shapes. To do that, we calculate the ratio of the average laser beam powers measured at the output plane of the CFEM and before the SLM with the help of a power meter (GENTEC Tuner). Our results confirm that, apart from common energy losses, the efficiency of our method is pattern-size dependent. For instance, when the encoded pattern utilizes less than 10% of the active area (about 133 mm$^2$) of the SLM, the classical CGH-based illumination method is more efficient than ours (for instance for encoded pattern dimensions of about 8 mm$^2$, we get 18% efficiency against 23% of the classical method). In contrast, when encoded patterns fill out areas greater than 25% of the SLM display area which is perhaps a more useful scenario from an energy point of view, the efficiency of our method exceeds that of the classical one (for instance for encoded pattern dimensions of about 60 mm$^2$, we get 32 % efficiency of our method against 23% of the classical one).

Another important point of the CIM is its potential to perform wide-field illumination and nonlinear excitation of biological samples without damage of optical components. On this aspect, the critical component of the ODM is clearly the microscope objective $MO_1$ that is working under low but focusing incident light conditions. The use of long-distance focal lengths certainly allows light energy inside the microcopy objective to be axially distributed along the associated Rayleigh range, decreasing in this way potential optical damage in comparison with the use of shorter focal length lenses. We measured the threshold average beam power at which nonlinear effects appear in the microscope objective. This value was about 20 mW, which turned out to be relative far from the typical average powers (4-10 mW) that we use in our experiments.

Finally we discuss on the dispersion compensation capabilities of the CIM. It is apparent that because of the joint effects of material dispersion, propagation time difference, angular dispersion, as well as general first and high order aberrations introduced by optical elements of the optical setup, the ultrashort pulse can be both spatial and temporal stretched at the sample plane. However, as the proposed illumination method is experimentally implemented through imaging-based optical systems, the distortions in the spectral phase of the ultrashort pulse can be pre-compensated (at least at first order) with a common dispersion compensation module. In our case, we employed a commercially available module (Dazzler) installed in the amplification stage of our femtosecond laser. For instance, temporal pulse stretching is mainly mitigated with a proper combination of second and third order terms if we look at the maximum second harmonic signal of a BBO crystal at the sample plane. Additionally, when necessary, spatial aberrations of the illumination patterns at the sample plane can also be partially compensated by encoding a suited phase mask (makes up of a set of weighted aberration Zernike polynomials) together with phase element $\alpha(x, y)$ given in Eq. (4). Note that, the lack of significant dispersion contributions at the sample plane can be regarded as an advantage of our proposal in comparison with other illumination strategies e.g., CGHs, where temporal and spatial broadening of the ultrashort pulse at the reconstruction plane are usually not trivial to compensate [20].

## Conclusions

In this manuscript we have investigated a wide-field CIM for linear and nonlinear microscopy applications. This method allows not only simultaneous but also full control of the amplitude and phase of the complex field over different ROIs of a sample. All this is basically accomplished by means of a single phase element implemented onto a phase-only SLM. For instance, uniform irradiance illumination of an arbitrary-shape ROI of about 60x60 μm$^2$ was carried without coherent noise, with an intensity profile accuracy similar to that of the collimated laser beam at the SLM plane. Here, it should be noted that using CGH-based illumination methods we cannot generate controllable excitation patterns having different energy level distributions, but even less simultaneous and independent amplitude and phase control over reconstructed holograms. This mainly happens because Fresnel holograms can only approximate the amplitude (or phase) of the complex field at the reconstruction plane.

To support our proposal, we have discussed both the theory underlying the present CIM as well as its experimental performance, including zero order behavior, dispersion compensation capabilities or efficiency. From an experimental point of view, we have shown that CIM can be used to produce wide-field SHG signal from arbitrary-shape ROIs. We believe that the ability of the CIM to simultaneous excite spatially-extended ROIs and also to compensate for optical aberrations at the sample plane should positively contribute to speed up acquisition time and improve the spatial resolution of acquired images in nonlinear microscopy applications. To this end, a proper combination of the present CIM with high-sensitivity cameras is expected. Finally, we would like to comment that CIM could also be also utilized to explore 3D (instead of 2D) excitation of biological samples. In this case, main challenge should be to manage light dispersion at the tissue rather than encoding multilayer irradiance patterns at the same time.

## Acknowledgments

The authors are very grateful to the SCIC of the Universitat Jaume I for the use of the femtosecond laser. This study was funded by Generalitat Valenciana (AICO/2016/036, PROMETEO 2016-079); Universitat Jaume I (UJI) (UJIB2016-19) & Ministerio de Economía y Competitividad (MINECO) (FIS2016-75618-R).

# References

1. Kim, C. K. *et al.* Simultaneous fast measurement of circuit dynamics at multiple sites across the mammalian brain. *Nat. Methods* **13**, 325–328 (2016).
2. Ducros, M., Houssen, Y. G., Bradley, J., Sars, V. de & Charpak, S. Encoded multisite two-photon microscopy. *Proc. Natl. Acad. Sci.* **110**, 13138–13143 (2013).
3. Butko, M. T. *et al.* Simultaneous multiple-excitation multiphoton microscopy yields increased imaging sensitivity and specificity. *BMC Biotechnol.* **11**, 20 (2011).
4. Ingaramo, M. *et al.* Two-photon excitation improves multifocal structured illumination microscopy in thick scattering tissue. *Proc. Natl. Acad. Sci.* **111**, 5254–5259 (2014).
5. Egner, A. & Hell, S. W. Time multiplexing and parallelization in multifocal multiphoton microscopy. *J. Opt. Soc. Am. A* **17**, 1192 (2000).
6. Leveque-Fort, S., Fontaine-Aupart, M.-P., Roger, G. & Georges, P. Fluorescence lifetime imaging with a multifocal two-photon microscope. in *Proceedings of SPIE* (eds. Periasamy, A. & So, P. T. C.) **29**, 99–107 (International Society for Optics and Photonics, 2004).
7. Papagiakoumou, E., de Sars, V., Oron, D. & Emiliani, V. Patterned two-photon illumination by spatiotemporal shaping of ultrashort pulses. *Opt. Express* **16**, 22039–22047 (2008).
8. Cheng, L.-C. *et al.* Spatiotemporal focusing-based widefield multiphoton microscopy for fast optical sectioning. *Opt. Express* **20**, 8939 (2012).
9. Peterson, M. D. *et al.* Second harmonic generation imaging with a kHz amplifier [Invited]. *Opt. Mater. Express* **1**, 57 (2011).
10. Pastirk, I., Cruz, J. M. Dela, Walowicz, K. A., Lozovoy, V. V. & Dantus, M. Selective two-photon microscopy with shaped femtosecond pulses. *Opt. Express* **11**, 1695 (2003).
11. Maurer, C., Jesacher, A., Bernet, S. & Ritsch-Marte, M. What spatial light modulators can do for optical microscopy. *Laser Photon. Rev.* **5**, 81–101 (2011).
12. Nikolenko, V., Peterka, D. S., Araya, R., Woodruff, A. & Yuste, R. Spatial Light Modulator Microscopy. *Cold Spring Harb. Protoc.* **2013**, 1132–1141 (2013).
13. Bovetti, S. & Fellin, T. Optical dissection of brain circuits with patterned illumination through the phase modulation of light. *J. Neurosci. Methods* **241**, 66–77 (2015).
14. Hertzberg, Y., Naor, O., Volovick, A. & Shoham, S. Towards multifocal ultrasonic neural stimulation: pattern generation algorithms. *J. Neural Eng.* **7**, 056002 (2010).
15. Mendoza-Yero, O., Mínguez-Vega, G. & Lancis, J. Encoding complex fields by using a phase-only optical element. *Opt. Lett.* **39**, 1740 (2014).
16. Mendoza-Yero, O. *et al.* Diffraction-Based Phase Calibration of Spatial Light Modulators With Binary Phase Fresnel Lenses. *J. Disp. Technol.* **12**, 1027–1032 (2016).
17. Persson, M., Engström, D. & Goksör, M. Reducing the effect of pixel crosstalk in phase only spatial light modulators. *Opt. Express* **20**, 22334 (2012).
18. Yamaguchi, I. & Zhang, T. Phase-shifting digital holography. *Opt. Lett.* **22**, 1268 (1997).
19. R. W., G. & W. O., S. A Practical Algorithm for the Determination of Phase from Image and Diffraction Plane Pictures. *Optik (Stuttg).* **35**, 237–246 (1972).
20. Martínez-León, L. *et al.* Spatial-chirp compensation in dynamical holograms reconstructed with ultrafast lasers. *Appl. Phys. Lett.* **94**, 011104 (2009).